\documentstyle[12pt]{article}

\catcode`\@=11 \@addtoreset{equation}{section}

\catcode`\@=12

\begin{document}

\title{\Large\bf Studies on the Weak Charges of Particles}
\author{Zhiqiang Shi \thanks{Present address: Residence 10-2-7,
Shaanxi Normal University, Xi'an 710062, P. R. China.
 E-mail address: zqshi@snnu.edu.cn}\\
\small\it{Faculty of Science, Xi'an Jiaotong University, Xi'an 710049, P. R. China}}

\date{}

\maketitle
\begin{abstract}
In order to include massive neutrino theoretically, a new concept
of the weak charges of the particles, fundamental fermions,
intermediate bosons and hadrons, is introduced. A new conservation
law of the weak charge is first reported. According to the
chirality of the weak charge the origin of parity nonconservation
in the weak interactions is reasonably explained. According to the
symmetry of the weak charge, an extension of the standard model is
proposed. In this scenario, all the three generations of neutrinos
are massive Dirac particles; both the right-handed neutrinos and
the right-handed neutral baryons have zero weak charge, and so
they do not take part in the weak interactions. \\[0.4cm] PACS
numbers: 12.60.Cn, 11.30.Er, 14.60.St, 13.30.-j
\end{abstract}

\hskip\parindent

In spite of its tremendous successes when confronted with
experiment, the standard model (SM) has some shortcomings and
leaves unanswered many fundamental questions. Perhaps one of the
important outstanding problems amongst them is the neutrino mass,
which would have a significant impact on astrophysics, cosmology
and particle physics. In the SM neutrinos are exactly massless.
However, theoretically here is no compelling reason for massless
neutrinos, and there exists already a vast literature on the model
of massive neutrinos [1]. Many extensions of the SM contain
neutrinos with Majorana or Dirac masses. Experimentally there is
now a host of evidence for neutrino oscillation [2], and that is
most naturally explained if neutrinos are massive and mix with one
another. Especially, the recent result of Super-Kamiokande
collaboration [3] gave the mass squared difference $\Delta
m^{2}=5\times 10^{-4}-6\times 10^{-3}$ eV$^2$ and the mixing angle
${\rm sin}^{2} 2\theta >0.82$. Therefore, some physicists believe
that neutrinos might indeed have a very small mass [4]. If
neutrinos have non-zero mass, it will radically alter our
understanding of the violation of parity conservation law in the
weak interactions and implies physics beyond the SM of particle
physics [5]. The massless neutrinos are inevitable outcome of the
parity nonconservation in the weak interactions. Therefor, in
order to include massive neutrino theoretically, a possible
suggestion is studying the root cause of the parity violation in
the weak interactions. In fact, as early as 1982 Professor Yang
emphasized again that a more fundamental origin of the violation
of discrete symmetry should be investigated [6].

In this paper we will be forced to propose the concept of the weak
charges of the particles in order to study the above issues. The
weak charges of the fundamental fermions are introduced by analogy
between the interaction Lagrangian for neutral weak currents and
the electromagnetic currents, the weak charges of the intermediate
bosons by analogy between the Lagrangian terms describing the
self-interaction of gauge fields. The weak charges of some baryons
and mesons are calculated. The question of parity nonconservation
in the weak interactions is explained by the chirality of the weak
charge. The conservation law of the total weak charge is tested in
all interactions and the conservation law of the chiral weak
charge, in the weak interactions. In the last section the SM is
extended to include the right-handed neutrinos.

\section{The weak charges of particles}
\subsection{The weak charges of the fundamental fermions}
\hskip\parindent
The existence of the three generations of the
fundamental fermions, leptons and quarks, has been established in
the SM. We shall confine ourselves to considering only one of
them, e.g., first generation because they are absolutely alike.
According to the SM we choose the group SU(2)$\times$U(1) as the
gauge group, and the left-handed (LH) fermions as the
SU(2)-doublets and the right-handed (RH) fermions as the
SU(2)-singlets,
$$\ell_{_L}=\left(\begin{array}{c}
\nu_{_L}\\e_{_L}
\end{array}\right),\quad q_{_L}=\left(\begin{array}{c}u_{_L}\\
d_{_L}\end{array} \right),\quad e_{_R},u_{_R},d_{_R}.\eqno{(1)}$$
The Lagrangian of the model contains the terms describing the
electromagnetic and the weak interactions. The Lagrangian for the
electromagnetic interactions of the fermions is
 $$ {\cal L}_{em}=
-{\displaystyle \frac{g_1g_2}{\sqrt{g_1^2+g_2^2}}}\: {\overline e}\:\gamma_\mu A_{\mu}e
+{\displaystyle \frac{2}{3}\frac{g_1g_2}{\sqrt{g_1^2+g_2^2}}}\: {\overline u}\:\gamma_\mu
A_{\mu}u -{\displaystyle \frac{1}{3}\frac{g_1g_2}{\sqrt{g_1^2+g_2^2}}}\: {\overline
d}\:\gamma_\mu A_{\mu}d,\eqno{(2)}$$ where $A_{\mu}$ denotes the electromagnetic field.
We see that the rationalized electric charge is
$$e=\frac{g_1g_2}{\sqrt{g_1^2+g_2^2}},\eqno{(3)}$$
where
$g_1$ and $g_2$ are the coupling constants corresponding to the
groups U(1) and SU(2), respectively.

The most interesting property of the SM is the occurrence of neutral weak currents. A
number of consequences of this prediction have been subject to experimental checks. The
interaction Lagrangian for the neutral weak lepton current reads $${\cal
L}_{\ell}=\frac{1}{2}\sqrt{g_1^2+g_2^2} \:{\overline{\nu}}_{_L}\gamma_\mu Z_{\mu}\nu_{_L}
-{\displaystyle \frac{1}{2}\frac{g_2^2-g_1^2}{\sqrt{g_1^2+g_2^2}}}
\:{\overline{e}}_{_L}\gamma_\mu Z_{\mu}e_{_L} +{\displaystyle
\frac{g_{1}^{2}}{\sqrt{g_1^2+g_2^2}}} \:{\overline{e}}_{_R}\gamma_\mu
Z_{\mu}e_{_R},\eqno{(4)}$$ where $Z_{\mu}$ denotes the neutral intermediate vector boson
field. The interaction Lagrangian for the neutral weak quark current reads $$
\begin{array}{rl}
{\cal L}_{q}=&{\displaystyle \frac{1}{2}\frac{g_2^2-\frac{1}{3}g_1^2}
{\sqrt{g_1^2+g_2^{2}}}}\:{\overline u}_{_L}\gamma_\mu Z_{\mu}u_{_L} -{\displaystyle
\frac{2}{3}\frac{g_{1}^{2}}{\sqrt{g_1^2+g_2^2}}} \:{\overline u}_{_R}\gamma_\mu
Z_{\mu}u_{_R}\cr\noalign{\vskip2mm}
&-{\displaystyle\frac{1}{2}\frac{g_2^2+\frac{1}{3}g_1^2}
{\sqrt{g_1^2+g_2^2}}}\:{\overline d}_{_L}\gamma_\mu Z_{\mu}d_{_L} +{\displaystyle
\frac{1}{3}\frac{g_1^2}{\sqrt{g_1^2+g_2^2}}} \:{\overline d}_{_R}\gamma_\mu
Z_{\mu}d_{_R}.\end{array}\eqno{(5)}$$ In analogy with the Lagrangian (2) for the
electromagnetic interaction, from (4) and (5) we assume that each fundamental fermion
possesses its own weak charge $w$, which was first reported in Ref.[7], to wit $$
\begin{array}{ll}
w(\nu_{_L})={\displaystyle
\frac{1}{2}\sqrt{g_1^2+g_2^2}},\hspace{1.5cm} &
\cr\noalign{\vskip2mm} w(e_{_L})= {\displaystyle
-\frac{1}{2}\frac{g_2^2-g_1^2}{\sqrt{g_1^2+g_2^2}}},
&w(e_{_R})={\displaystyle
\frac{g_1^2}{\sqrt{g_1^2+g_2^2}}},\cr\noalign{\vskip2mm}
w(u_{_L})={\displaystyle \frac{1}{2}\frac{g_2^2-\frac{1}{3}g_1^2}
{\sqrt{g_1^2+g_2^2}}}, &w(u_{_R})={\displaystyle
-\frac{2}{3}\frac{g_1^2}{\sqrt{g_1^2+g_2^2}}},
\cr\noalign{\vskip2mm} w(d_{_L})={\displaystyle
-\frac{1}{2}\frac{g_2^2+\frac{1}{3}g_1^2} {\sqrt{g_1^2+g_2^2}}},
&w(d_{_R})={\displaystyle
\frac{1}{3}\frac{g_1^2}{\sqrt{g_1^2+g_2^2}}},\end{array}\eqno{(6)}$$
where $ w(f_{_L})$ is the weak charge of the LH fermion $f_{_L}$
and $w(f_{_R})$ that of the RH fermion $f_{_R}$, $f=\nu, e, u,$ or
$d$. Setting $$w_1=\frac{g_1^2}{\sqrt{g_1^2+g_2^2}},\quad
w_2=\frac{g_2^2}{\sqrt{g_1^2+g_2^2}},\eqno{(7)}$$ then
$$e=\sqrt{w_{1}w_2},\eqno{(8)}$$ and the expression (6) for the
weak charge can be written as $$\begin{array}{ll}
w(\nu_{_L})={\displaystyle \frac{1}{2}w_1+\frac{1}{2}w_2},
\hspace{1.5cm}& \cr\noalign{\vskip2mm} w(e_{_L})={\displaystyle
\frac{1}{2}w_1-\frac{1}{2}w_2},& w(e_{_R})={\displaystyle
\frac{1}{2}w_1+\frac{1}{2}w_1}, \cr\noalign{\vskip2mm}
w(u_{_L})={\displaystyle -\frac{1}{6}w_1+\frac{1}{2}w_2}, &
w(u_{_R})={\displaystyle -\frac{1}{6}w_1-\frac{1}{2}w_1},
\cr\noalign{\vskip2mm} w(d_{_L})={\displaystyle
-\frac{1}{6}w_1-\frac{1}{2}w_2}, & w(d_{_R})={\displaystyle
-\frac{1}{6}w_1+\frac{1}{2}w_1}.\end{array}\eqno{(9)}$$

\subsection{The weak charges of the intermediate bosons}
\hskip\parindent In the SM,the Lagrangian describing the
self-interaction of gauge fields is given by
$$
\begin{array}{ll}
{\cal L}_{gi}=&{\displaystyle\frac{g_2^4}{g_1^2+g_2^2}}\:(W_{\nu}^{+}W^{-}_\mu Z_\mu
Z_\nu-W_{\nu}^{+}W^{-}_\nu Z_\mu Z_\mu)\cr\noalign{\vskip2mm}
&+{\displaystyle\frac{g_1^{2}g_2^2}{g_1^2+g_2^2}}\: (W_{\nu}^{+}W^{-}_\mu A_\mu A_\nu-
W_{\nu}^{+}W^{-}_\nu A_\mu A_\mu)\cr\noalign{\vskip2mm} &+{\displaystyle
\frac{1}{2}}g_2^2\:(W_{\nu}^{+}W_{\mu}^{-} W^{+}_\nu
W^{-}_\mu-W_{\nu}^{+}W_{\mu}^{-}W^{-}_\nu W^{+}_\mu)\cr\noalign{\vskip2mm}
&+{\displaystyle \frac{g_1g_2^3}{g_1^2+g_2^2}}\:(W_{\nu}^{+}W^ {-}_\mu Z_\mu A_\nu
+W_{\nu}^{+}W^{-}_\mu A_\mu Z_\nu) \cr\noalign{\vskip2mm} &+2{\displaystyle
\frac{g_1g_2^3}{g_1^2+g_2^2}}\:W_{\nu}^{+}W^{-}_\nu Z_\mu A_\mu.\end{array}\eqno{(10)}$$
From the second term it is found that the electromagnetic interaction occur between the
electromagnetic field $A_\mu$ and the charged intermediate vector boson field $W^\pm$,
and $W^+$ and $W^-$ boson have the electric charge $e$ and $-e$, respectively. Comparing
the first term with the second term it is assumed that $W^+$ and $W^-$ boson possess the
weak charge $w(W^+)$ and $w(W^-)$, respectively, and $$ w(W^+)=\frac{g_2^2}{\sqrt{
g_1^2+g_2^2}}=w_2, \eqno{(11)}$$ $$ w(W^-)=-\frac{g_2^2}{\sqrt{ g_1^2+g_2^2}}=-w_2.
\eqno{(12)}$$ As indicated from formula (10), both the electromagnetic field $A_\mu$ and
the neutral intermediate vector boson field $Z_\mu$ have no self-interaction, hence the
photon has zero electric charge, and both it and $Z$ boson have zero weak charge.

\subsection{The weak charges of the hadrons}
\hskip\parindent The resultant weak charge of a system is the
algebraic sum of the weak charges of all particles, which are
involved in the system. The hadrons are made out of quarks. Hence,
the left-handed weak charge (LHWC) of a hadron is equal to the sum
of the LHWC of the individual quarks, and the right-handed weak
charge (RHWC) of a hadron the sum of the RHWC of the quarks which
are involved in the hadron. Considering the operation $CP$ is
conserved (so far, $CP$ violation has only been observed in the
$K^0\overline K{^0}$ system), and the electric charges of a
particle and an antiparticle are always opposite in the signs, we
assume that the LHWC of a fundamental fermion $f$ and the RHWC of
an antifermion $\overline f$ have equal magnitude but opposite
sign, i.e., $$w(f_{_L})=-w({\overline f}_{_R})\quad or \quad
w(f_{_R})=-w({\overline f}_{_L}).\eqno{(13)}$$ Thus the weak
charges of some baryons and mesons can be calculated, and are
listed in Table \ref{tab-1} and \ref{tab-2}, respectively. The
meson has zero spin, and there exists not the LH or RH meson,
therefor the LHWC of a meson should be regarded as the weak charge
in a LH system of coordinates, and the RHWC of a meson in a RH
system. Since the quark combination state of $\pi^0$ meson is $(u
\overline u- d {\overline d})/\sqrt{2}$, the weak charges of two
kinds of quark-antiquark pairs, $u \overline u$ and $ d {\overline
d} $, are calculated in the Table \ref{tab-2}. The weak charge of
$\eta$-meson is also calculated with the same way. From Table
\ref{tab-1} and \ref{tab-2}, it is easy to see that the LHWC of a
hadron is generally different from its RHWC. But, the RHWC of a
neutral baryon is equal to zero, and the LHWC of a charged meson
is identical with its RHWC.

\begin{table*}[p]
\caption{The weak charges of baryons}\label{tab-1}
\begin{center}
\renewcommand{\arraystretch}{1.6}
\small
\tabcolsep 10pt
\begin{tabular}{cccccc}
\hline\hline Particles&$J^P$&Quark
Composition&LHWC&RHWC&$\gamma_w$\\ \hline $\rm P$& &$uud$& &\\
$\Lambda_0 ^+$& &$udc$& &\\ $\Sigma
^+$&$\frac{1}{2}^+$&$uus$&$-\frac{1}{2}(w_{1}-w_2) $&$-w_1$&$+1$\\
$\Sigma_0 ^+$& &$udc$& &\\ $\Xi_0 ^+$& &$usc$& &\\ \hline $\rm n$&
&$udd$& &\\ $\Lambda$& &$uds$& &\\ $\Sigma
^0$&$\frac{1}{2}^+$&$uds$&$-\frac{1}{2}(w_{1}+w_2) $&$0$&$-1$\\
$\Sigma_0 ^0$& &$ddc$& &\\ $\Xi ^0$& &$uss$& &\\ \hline $\Sigma
^-$&$\frac{1}{2}^+$&$dds$& &\\ $\Xi
^-$&$\frac{1}{2}^+$&$dss$&$-\frac{1}{2}(w_{1}+3w_2) $&$w_1$&$-3$\\
$\Omega ^-$&$\frac{3}{2}^+$&$sss$& &\\ \hline
$\Sigma_0^{++}$&$\frac{1}{2}^+$&$uuc$&$-\frac{1}{2}(w_{1}-3w_2)$&
$-2w_1$&$+3$\\
$\Delta^{++}$&$\frac{3}{2}^+$&$uuu$&$-\frac{1}{2}(w_{1}-3w_2)$&$-2w_1$&$+3$\\
\hline\hline
\end{tabular}\end{center}\end{table*}

\begin{table*}[p]
\caption{The weak charges of mesons}\label{tab-2}
\begin{center}
\renewcommand{\arraystretch}{1.6}
\small
\tabcolsep 10pt
\begin{tabular}{cccccc}
\hline\hline Particles&$J^P$&Quark
composition&LHWC&RHWC&$\gamma_w$\\ \hline
$\pi^+$&$0^-$&$u\overline d$&$-\frac{1}{2}(w_{1}-w_2) $&$-
\frac{1}{2}(w_{1}-w_2)$&0\\ $K ^+$&$0^-$&$u\overline
s$&$-\frac{1}{2}(w_{1}-w_2) $&$- \frac{1}{2}(w_{1}-w_2)$&0\\ $D
^+$&$0^-$&$c\overline d$&$-\frac{1}{2}(w_{1}-w_2) $&$-
\frac{1}{2}(w_{1}-w_2)$&0\\ $B ^+$&$0^-$&$u\overline
b$&$-\frac{1}{2}(w_{1}-w_2) $&$- \frac{1}{2}(w_{1}-w_2)$&0\\
\hline $\pi ^-$&$0^-$&${\overline u}d$&$\frac{1}{2}(w_{1}-w_2) $
&$\frac{1}{2}(w_{1}-w_2)$&0\\ $K ^-$&$0^-$&${\overline
u}s$&$\frac{1}{2}(w_{1}-w_2) $ &$\frac{1}{2}(w_{1}-w_2)$&0\\ $D
^-$&$0^-$&${\overline c}d$&$\frac{1}{2}(w_{1}-w_2) $
&$\frac{1}{2}(w_{1}-w_2)$&0\\ $B ^-$&$0^-$&${\overline
u}b$&$\frac{1}{2}(w_{1}-w_2) $ &$\frac{1}{2}(w_{1}-w_2)$&0\\
\hline $\pi ^0$&$0^-$&$u\overline u$&$\frac{1}{2}(w_{1}+w_2) $&$-
\frac{1}{2}(w_{1}+w_2)$&$+2$\\ & &$d\overline
d$&$-\frac{1}{2}(w_{1}+w_2)$&$\frac{1}{2}(w_{1}+w_2)$&$-2$\\
\hline & &$u\overline u$&$\frac{1}{2}(w_{1}+w_2) $&$-
\frac{1}{2}(w_{1}+w_2)$&$+2$\\ $\eta$&$0^-$&$d\overline
d$&$-\frac{1}{2}(w_{1}+w_2) $&$\frac{1}{2}(w_{1}+w_2)$&$-2$\\ &
&$s\overline s$&$-\frac{1}{2}(w_{1}+w_2)
$&$\frac{1}{2}(w_{1}+w_2)$&$-2$\\ \hline $K^0$&$0^-$&$d\overline
s$&$-\frac{1}{2}(w_{1}+w_2) $&$\frac{1}{2}(w_{1}+ w_2)$&$-2$\\
$\overline K{^0}$&$0^-$&${\overline d}s$&$-\frac{1}{2}(w_{1}+w_2)
$& $\frac{1}{2}(w_{1}+w_2)$&$-2$\\ \hline $D^0$&$0^-$&$c\overline
u$&$\frac{1}{2}(w_{1}+w_2) $&$-\frac{1}{2}(w_{1}+ w_2)$&$+2$\\
$\overline D{^0}$&$0^-$&${\overline c}u$&$\frac{1}{2}(w_{1}+w_2)
$& $-\frac{1}{2}(w_{1}+w_2)$&$+2$\\ \hline $B^0$&$0^-$&$d\overline
b$&$-\frac{1}{2}(w_{1}+w_2) $&$\frac{1}{2}(w_{1}+ w_2)$&$-2$\\
$\overline B{^0}$&$0^-$&${\overline d}b$&$-\frac{1}{2}(w_{1}+w_2)
$& $\frac{1}{2}(w_{1}+w_2)$&$-2$\\ \hline\hline
\end{tabular}\end{center}\end{table*}

\section{Question of parity nonconservation}
\hskip\parindent As mentioned above, the LHWC and the RHWC of a
fundamental fermion are different. The chiral value of the weak
charge is defined as $$\gamma_w =\frac{w(f_{_L})-
w(f_{_R})}{w(\nu_{_L})- w(\nu_{_R})}.\eqno{(14)}$$ That is to say,
that $\gamma_w = +1$ for neutrino and $u$ quark, $\gamma_w = -1$
for electron and $d$ quark, and an antiparticle has the same
$\gamma_w$ as a particle. The chiral value of a hadron is equal to
the sum of the chiral values of all quarks, which are involved in
the hadron. The chiral value of some baryons and mesons are listed
in Table \ref{tab-1} and \ref{tab-2}, respectively. It is found
that the baryons have odd chiral values, i.e., $\gamma_w =\pm 1$
or $\pm 3$, and the mesons have even chiral values, i.e.,
$\gamma_w =0$ or $\pm 2$. Particularly, the charged mesons have
zero chiral values.

For all weak interactions involving fermions, the chirality of every particle has been
analyzed in detail. In the neutron-decay, for example, since the RH neutron has zero weak
charge (see table \ref{tab-1}), it does not take part in weak interaction, and the
neutron must be the LH particle. The experiments have proved [8] that the electron is the
LH particle, and the antineutrino is the RH in the $\beta$-decays, and then the proton
must be also the LH particles in view of the conservation law of angular momentum, i.e.,
$$ n_{_L}\longrightarrow p_{_L}+ e_{_L}+ \overline \nu _{_R}, \eqno{(15)}$$ Obviously,
the parity is not conserved in this process.

The pion and muon decay schemes are
$$\pi^+\longrightarrow\mu_{_R}^{+}+\nu_{\mu{_L}}\eqno{(16)}$$ and
$$\mu_{_R}^+\longrightarrow e_{_R}^{+}+\nu_{e{_L}}+\overline
\nu_{\mu{_R}}. \eqno{(17)}$$ Since the pion has spin zero the muon and the neutrino must
have antiparallel spin vectors. The muon and the neutrino have parallel momentum vectors
in the laboratory coordinates, and the neutrino is LH particle, so that the $\mu^+$ must
be RH. In the subsequent muon decay, the experiment has shown the positron $e^+$ to be
the RH particle [9]. Clearly, all of fermions are chiral and the parity is naturally
violated in the two processes.

For the purely meson processes, the analysis of angular momentum
can not be employed, however, the change of parity can be
explained by the change of the chiral value of the weak charge. If
the sum total of chiral value of a system is constant before and
after decay, i.e., $\Delta\gamma_w=0$, then the parity must be
conserved; conversely, if $\Delta\gamma_w\not=0$, then the parity
must be not conserved. For example, the value of $\gamma_w$
involved in the $K$-decay are as follows: $$ K^+\rightarrow
\pi^{+}+\pi^{+}+\pi^-,\hspace{1.0cm} K^0\rightarrow
\pi^{+}+\pi^{-}+\pi^0.\eqno{(18)}$$
\hspace{2.8cm}$\gamma_w$\hspace{0.9cm}0\hspace{1.1cm}0\hspace{0.8cm}0\hspace{0.8cm}0
\hspace{1.3cm}-2\hspace{0.9cm}0\hspace{0.8cm}0\hspace{0.8cm}-2\\[0.2cm]
The total chiral value of every system is constant on the two
sides of the equation, i.e., $\Delta\gamma_w=0$, therefor the
parity is conserved. In the processes $$K^+\longrightarrow
\pi^{+}+\pi^0,\hspace{1.5cm} K^0\longrightarrow
\pi^{+}+\pi^{-},\eqno{(19)}$$
\hspace{3.5cm}$\gamma_w$\hspace{0.8cm}0\hspace{1.2cm}0\hspace{0.8cm}-2
\hspace{1.6cm}-2\hspace{1.2cm}0\hspace{0.8cm}0\\[0.2cm]
$\Delta\gamma_w\not=0$, so that the parity is not conserved.

In brief, the weak charges of particles cause them to exert weak
interactions on one another, like electric charge. Therefor the
left-right asymmetry of the weak charge certainly leads to that
the Hamiltonian ${\bf H}$ describing weak interactions will not be
invariant under inversion of space coordinates, and parity
conservation law must be violated. Because the electric charges
and the colour charges of fundamental fermions do not exhibit any
chirality, the parity is conserved in the strong and
electromagnetic interactions. T. D. Lee and C. N. Yang said:
``Should it further turn out that the two-component theory of the
neutrino described above is correct, one would have a natural
understanding of the violation of parity conservation in processes
involving the neutrino. An understanding of the $\theta - \tau$
puzzle presents now a problem on a new level because no neutrinos
are involved in the decay of $K_{\pi 2}$ and $K_{\pi 3}$. Perhaps
this means that a more fundamental theoretical question should be
investigated: the origin of all weak interactions.'' [10] Now, by
introducing the concept of weak charge, this problem can be
reasonably and intuitively explained.

\section{The conservation law of the weak charge}
\hskip\parindent For all weak interactions, the change of the weak
charge of every system has been calculated in detail, and the
conservation law of the chiral weak charge is discovered. To
understand this let us consider, for example, the $\beta$-decay.
In the neutron-decay (15), according to the formula (9) and Table
 \ref{tab-1}, we get
 $$ w(n_{_L})=w(p_{_L})+w(e_{_L})+w(\overline
\nu_{_R})\eqno{(20)}$$ That is to say, the chiral weak charge of
the system is constant before and after decay.

The conservation law of the chiral weak charge can be confirmed in terms of the Feynman
diagrams. The interaction Lagrangian for the charged weak lepton currents is given by
$${\cal L}_{\ell \:W}= \frac{1}{\sqrt{2}}\:g_2\:{\overline e_{_L}}\gamma_\mu
W_\mu^+\nu_{_L} +\frac{1}{\sqrt{2}}\:g_2\:{\overline \nu_{_L}}\gamma_\mu W_\mu^-e_{_L}.
\eqno{(21)}$$ The Feynman diagrams corresponding to the interactions are shown in Fig. 1.
\begin{center}
\unitlength=0.7mm
\begin{picture}(100,35)
\put(0,0){\vector(1,1){7.5}}
\put(7.5,7.5){\line(1,1){7.5}}
\put(15,15){\vector(-1,1){7.5}}
\put(7.5,22.5){\line(-1,1){7.5}}
\multiput(15,15)(3,0){5}{\line(1,0){2}}
\put(30,15){\vector(1,0){2}}
\multiput(33,15)(3,0){4}{\line(1,0){2}}
\put(65,0){\vector(1,1){7.5}}
\put(72.5,7.5){\line(1,1){7.5}}
\put(80,15){\vector(-1,1){7.5}}
\put(72.5,22.5){\line(-1,1){7.5}}
\multiput(80,15)(3,0){5}{\line(1,0){2}}
\put(95,15){\vector(1,0){2}}
\multiput(98,15)(3,0){4}{\line(1,0){2}}
\put(7,3){$\nu_{_L}$}
\put(8,24){$e^-_{_L}$}
\put(35,17){$W^+$}
\put(72,3){$e^-_{_L}$}
\put(73,24){$\nu_{_L}$}
\put(100,17){$W^-$}
\end {picture}
\end{center}
\begin{center}
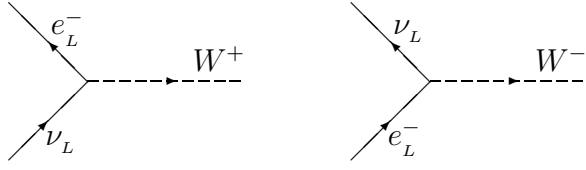
FIG. 1. Diagram for the weak interaction of an electron with a neutrino.
\end{center}
The interaction Lagrangian for the charged weak quark currents is given by $$ {\cal L}_{q
W}= \frac{1}{\sqrt{2}}\:g_2 \:{\overline d_{_L}}\gamma_\mu W_\mu^+u_{_L}
+\frac{1}{\sqrt{2}}\:g_2\: {\overline u_{_L}}\gamma_\mu W_\mu^-d_{_L}.\eqno{(22)}$$ The
Feynman diagrams corresponding to the interactions are shown in Fig. 2.
\begin{center}
\unitlength=0.7mm
\begin{picture}(100,35)
\put(0,0){\vector(1,1){7.5}}
\put(7.5,7.5){\line(1,1){7.5}}
\put(15,15){\vector(-1,1){7.5}}
\put(7.5,22.5){\line(-1,1){7.5}}
\multiput(15,15)(3,0){6}{\line(1,0){2}}
\put(30,15){\vector(1,0){2}}
\multiput(33,15)(3,0){4}{\line(1,0){2}}
\put(65,0){\vector(1,1){7.5}}
\put(72.5,7.5){\line(1,1){7.5}}
\put(80,15){\vector(-1,1){7.5}}
\put(72.5,22.5){\line(-1,1){7.5}}
\multiput(80,15)(3,0){6}{\line(1,0){2}}
\put(95,15){\vector(1,0){2}}
\multiput(98,15)(3,0){4}{\line(1,0){2}}
\put(8,3){$u_{_L}$}
\put(8,24){$d_{_L}$}
\put(35,17){$W^+$}
\put(72,3){$d_{_L}$}
\put(73,24){$u_{_L}$}
\put(100,17){$W^-$}
\end {picture}
\end{center}
\begin{center}
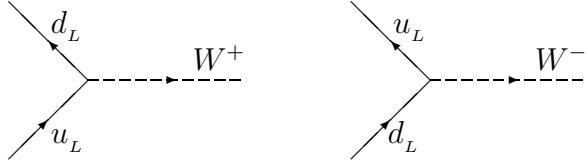
FIG. 2. Diagram for the weak interaction of a $u$ quark with a $d$ quark.
\end{center}
It is obvious from Fig.1 and Fig.2 that the chiral weak charge is
always conserved at every vertex of Feynman diagrams, like the
electric charge, angular momentum, etc.

For the purely meson weak interactions, if the intrinsic parity of
a system is conserved the LHWC and the RHWC of the system will be
conserved respectively, such as processes (18). Otherwise, if the
intrinsic parity of a system is not conserved, the LHWC and the
RHWC of the system will be not conserved, respectively, such as
the processes (19).

It is surprisedly found that, however, the total net weak charge of any system, the
sum of the LHWC and the RHWC, is always conserved in all weak interactions, although
the LHWC and the RHWC of the system may be not conserved, respectively. For strong and
electromagnetic interaction processes the conservation law of the total weak charge
has also been inspected one by one, and found this to be so.

Lastly, we can draw the diagram of the $\pi^+$ decay, as shown in
Fig.3, which involving an intermediate $p\overline n$ state. At
strong interaction vertex A, the total weak charge is conserved.
At weak interaction vertex B and C the chiral weak charge is also
conserved, respectively.
\begin{center}
\unitlength=1.0mm
\begin{picture}(80,33)
\put(0,15){\vector(1,0){10.5}}
\put(10.5,15){\line(1,0){7.5}}
\put(18,15){\circle*{1}}
\put(25,15){\circle{20}}
\multiput(25,3)(0,3){8}{\line(0,1){2}}
\put(32,15){\circle*{1}}
\multiput(32,15)(3,0){5}{\line(1,0){2}}
\put(47,15){\vector(1,0){2}}
\multiput(50,15)(3,0){4}{\line(1,0){2}}
\put(61,15){\circle*{1}}
\put(61,15){\vector(1,1){7.5}}
\put(68.5,22.5){\line(1,1){7.5}}
\put(61,15){\line(1,-1){7.5}}
\put(76,0){\vector(-1,1){7.5}}
\put(5,17){$\pi^+$}
\put(15,16){$A$}
\put(18,21){$p$}
\put(16,9){$ \overline n$}
\put(30,21){$u_{_L}$}
\put(32,8){$\overline d_{_R}$}
\put(32,16){$B$}
\put(47,17){$W^+$}
\put(59,16){$C$}
\put(64,3){$\mu^{+}_{_R}$}
\put(62,24){$\nu_{\mu{_L}}$}
\end {picture}
\end{center}
\begin{center}
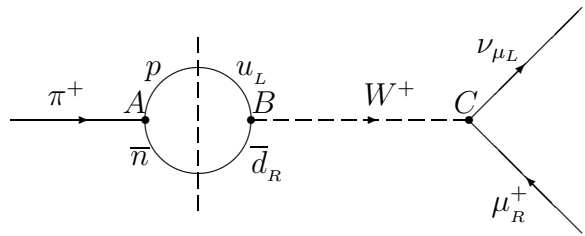
Fig. 3. Diagram for the decay of $\pi^+$ meson.
\end{center}

\section{The extension of the standard model}
\hskip\parindent As seen from formula (9), there exist the symmetries of the weak charge
among the fundamental fermions, and between the LH and the RH fermions. In accordance
with this symmetry, it is natural to assume that the RH neutrinos might be present in the
universe and possess weak charge $$ w(\nu_{_R})=\frac{1}{2}w_1-\frac{1}{2}w_1=0.
\eqno{(23)}$$ This implies that the neutrinos are massive. In the SM, the neutrino has
only the LH states, the antineutrino has only the RH states, and their masses must be
zero. Now, one can extend the SM gauge group to include the RH neutrino fields $\nu_{_R}$
as SU(2)-singlets. Obviously, the weak isospin and the supercharge of the RH neutrino are
both zero, so that the RH neutrino fields $\nu_{_R}$ are not present in interaction
Lagrangian for the neutral and the charged weak lepton currents. We also extend the SM to
include the Higgs scalar fields $\phi^\prime$ as SU(2)-doublet, which is charge conjugate
of the standard Higgs doublet $\phi$, i.e., $\phi^{\prime}=-i\tau_2 \phi^\ast$.
Dirac-type mass terms are generated by the Yukawa-type interaction of the lepton doublets
or singlets with Higgs scalar fields. The mass terms thus generated can contain, in
addition to the ordinary mass term of the charged lepton, the mass term of neutrino,
which is similar to that of $u$ quark. In this scenario neutrinos are treated on an equal
footing with the other fermions of the theory, and there exists a complete analogy
between the weak interaction of leptons and quarks. Both the RH neutrino and the LH
antineutrino have zero weak charges, so that they do not take part in weak interaction
and are termed the dark particles. B. Pontecorve [11] called them the `sterile'
neutrinos, which with definite mass are Majorana particles, not Dirac particles. With the
dark-massive Dirac neutrinos it is possible to explain simultaneously the solar neutrino
deficit [12], the atmospheric neutrino anomaly [13], the LSND data [14], and the hot dark
matter problem of the universe [15].

In conclusion the standard model has been extended to include the right-handed neutrinos.
Similar theory has been discussed before [16], however, there exist some outstanding
characteristics in this new scenario:

(1) Not only RH neutrinos, but also RH neutral baryons have zero weak charges, and they
do not take part in weak interactions. Therefor the lifetime of right-handed polarization
neutral baryons should be greater than that of left-handed polarization neutral baryons.
This problem will be discussed elsewhere. An experiment with a lifetime difference
between RH and LH polarization baryons, for example neutrons, needs to be performed. The
report on the experiment has not yet been discovered from the literature.

(2) The concept of electric charge is familiar to most people, while the weak charge is
something new. In Fermi weak interaction theory, coupling constant $G$ is called weak
charge in analogy with the current-current interaction in QED. In the SM, coupling
constant $g_2$ is called weak charge in analogy with the $V-A$ theory. Either of $G$ and
$g_2$ is a universal constant for all particles. However, the weak charges mentioned in
this paper are introduced by analogy between the Lagrangian for neutral weak currents and
the electromagnetic currents, and different particles have different weak charges. In
addition, it is more significant that the weak charges possess the symmetry and the
chirality.

(3) Comparing formula (8) with (9) one can see that the conservation law of the weak
charge is essentially different than that of electric charge. In the SM, the electric
charge is the only conserved quantum number after spontaneously broken symmetry.
Therefor, the weak charge added a new conserved quantum number to the extension model
proposed in this paper. The conservation of the weak charge should correspond to a
certain internal symmetry of the fundamental fermion, which is at present unknown,
implies some physics beyond the SM and needs to be investigated further.

(4) As seen from (9), the weak charge of a fundament fermion can be attributed to the
linear combination of two parts, an inherent weak charge and an intrinsic chiral weak
charge. The inherent weak charge of a lepton is $\frac{1}{2}w_1$ and one of a quark
$-\frac{1}{2}w_1$. Because each quark has three kinds of colour, a coloured quark has the
inherent weak charge of $-\frac{1}{6}w_1$. The intrinsic chiral weak charge of the LH
fermion is $\pm\frac{1}{2}w_2$, and one of the RH fermion $\pm\frac{1}{2}w_1$. Thus, the
weak charge of a fermion consists of two parts, and the quantum number of each part can
have values of $+\frac{1}{2}$ or $-\frac{1}{2}$, like spin or isospin. Since the charged
weak interaction is seen to be dominated by a coupling to left-handed fermions, the weak
charge $w_2$ interacts on the charged intermediate boson $W^\pm$ and the weak charge
$w_1$, on the neutral intermediate boson $Z$. This property of the weak charge might hint
some information about the internal structure or certain symmetry of the fundamental
fermions.

\end{document}